\documentclass[prl,twocolumn,showpacs,preprintnumbers,superscriptaddress,amsmath,amssymb]{revtex4}

\usepackage{epsf,color,graphicx}

\usepackage[dvips]{epsfig}

\begin{document}

%\preprint{APS/123-QED}

\title{Optical spectroscopy of two-dimensional layered $(C_{6}H_{5}C_{2}H_{4}-NH_{3})_{2}-PbI_{4}$ perovskite}% Force line breaks with \\
\author{K. Gauthron}
\affiliation{Laboratoire de Photonique et nanostructures,
CNRS-UPR20, route de Nozay  91460 Marcoussis, France}
\author{J. S. Lauret}
\affiliation{Laboratoire de Photonique Quantique et Mol\'eculaire,
\'Ecole Normale Sup\'erieure de Cachan, 61 avenue du pr\'esident
Wilson 94235 Cachan cedex, France} \email{lauret@lpqm.ens-cachan.fr}
\author{L. Doyennette}
\affiliation{Laboratoire de Photonique Quantique et Mol\'eculaire,
\'Ecole Normale Sup\'erieure de Cachan, 61 avenue du pr\'esident
Wilson 94235 Cachan cedex, France}
\author{G. Lanty}
\affiliation{Laboratoire de Photonique Quantique et Mol\'eculaire,
\'Ecole Normale Sup\'erieure de Cachan, 61 avenue du pr\'esident
Wilson 94235 Cachan cedex, France}
\author{A. Al Choueiry}
\affiliation{Laboratoire de Photonique Quantique et Mol\'eculaire,
\'Ecole Normale Sup\'erieure de Cachan, 61 avenue du pr\'esident
Wilson 94235 Cachan cedex, France}
\author{S.J. Zhang}
\affiliation{Laboratoire de Photonique Quantique et Mol\'eculaire,
\'Ecole Normale Sup\'erieure de Cachan, 61 avenue du pr\'esident
Wilson 94235 Cachan cedex, France}
\author{A. Brehier}
\affiliation{Laboratoire de Photonique Quantique et Mol\'eculaire,
\'Ecole Normale Sup\'erieure de Cachan,
61 avenue du pr\'esident Wilson 94235 Cachan cedex, France}%
\author{L. Largeau}
\affiliation{Laboratoire de Photonique et nanostructures,
CNRS-UPR20, route de Nozay  91460 Marcoussis, France}
\author{O. Mauguin}
\affiliation{Laboratoire de Photonique et nanostructures,
CNRS-UPR20, route de Nozay  91460 Marcoussis, France}
\author{J. Bloch}
\affiliation{Laboratoire de Photonique et nanostructures,
CNRS-UPR20, route de Nozay  91460 Marcoussis, France}
\author{E. Deleporte}
\affiliation{Laboratoire de Photonique Quantique et Mol\'eculaire,
\'Ecole Normale Sup\'erieure de Cachan, 61 avenue du pr\'esident
Wilson 94235 Cachan cedex, France}

\date{\today}% It is always \today, today,

\begin{abstract}
We report on optical spectroscopy (photoluminescence and
photoluminescence excitation) on  two-dimensional self-organized
layers of $(C_{6}H_{5}C_{2}H_{4}-NH_{3})_{2}PbI_{4}$ perovskite.
Temperature and excitation power dependance of the optical spectra
gives a new insight into the excitonic and phononic properties of
this hybrid organic/inorganic semiconductor. In particular,
exciton-phonon interaction is found to be more than one order of
magnitude higher than in GaAs QWs. As a result, photoluminescence
emission lines have to be interpreted in the framework of a
polaron model.\end{abstract}

\pacs{Valid PACS appear here}% PACS, the Physics and Astronomy
\maketitle

Optical properties of soft materials have attracted much attention
for years thanks to their potential applications in
optoelectronics devices. In particular, these last years, an
increasing number of studies are dedicated on hybrid
organic-inorganic materials\cite{Mitzi01}, due to the possibility
of combining the properties both of inorganic materials (high
mobility, electrical pumping, band engineering) and of organic
materials (low cost technology, high luminescence quantum yield
at room temperature). In this context, organic-inorganic
perovkites, having a chemical formula
$\mathrm{(R-NH3)_{2}MX_{4}}$ where R is an organic chain, M is a
metal and X a halogen, represent a natural hybrid system. Such
perovskites present a great flexibility in their optical
properties: the spectral position of the excitonic transitions
can be tailored by substituting different halides X
\cite{Papavassiliou95,Parashkov2007}, the photoluminescence
efficiency can be tailored by changing the organic part R
\cite{Zhang2009}. This kind of perovskites has been studied both
in the framework of fundamental studies
\cite{Lanty2008a,Lanty2008b,Brehier2006,Wenus2006,Parashkov2007,
Fujisawa04b,Fujisawa04a,Shimizu04,Kondo98,Tanaka02,Papavassiliou95,Hattori96,Hong92a,Ishihara89,Ishihara90,Ishihara92,Xu91a,Hong92b,Xu91b,Kataoka93,Muljarov95,Era94}
and of applications in optoelectronic as the active material in a
distributed feedback laser for example \cite{Fujita98}. Recently
the strong coupling regime between the perovskite exciton and the
optical mode of a P\'erot-Fabry microcavity has been demonstrated
at room temperature in the UV range \cite{Lanty2008a} and in the
visible range \cite{Brehier2006, Lanty2008b}. The physical
properties of these new polaritons have now to be investigated.
In particular, the demonstration of polariton-polariton
interactions which lead to polariton scattering would be a
breakthrough for the physics of these new devices in the context
of the low threshold polariton lasers
\cite{Livrekavokin,bajoniPRL,christopoulos}. To evaluate these
possibilities, a good knowledge of the perovskite material
electronic properties is needed. As an example, the energy of the
phonons and the strength of the electron-phonon coupling will
indicate whether an efficient relaxation of perovskite polaritons
is conceivable. Additionally, the origin of the different
perosvkite luminescence lines has to be clarified to improve the
knowledge about the exciton  which couples to the cavity mode.

\begin{figure}
 \begin{center}
\includegraphics*[scale=0.6]{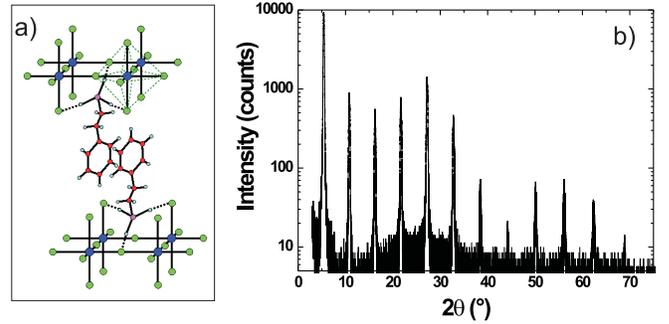} \caption[PL(T) ]{\textit{a) sketch of the perovskite structure. The blue points
at the center of the octahedrons represent the Pb atoms while I
atoms are displayed in green. The red symbols represents the
organic chains. b) X-ray diffraction spectrum of a 50 nm
$(C_{6}H_{5}C_{2}H_{4}-NH_{3})_{2}PbI_{4}$ layer.}}. \label{fig1}
\end{center}
\end{figure}
In this paper, we report on the optical properties of a particular
perovskite molecule, namely [bi-(phenethylammonium)
tetraiodoplumbate]: $(C_{6}H_{5}C_{2}H_{4}-NH_{3})_{2}-PbI_{4}$
(named PEPI), absorbing and emitting in the green part of the
visible range. Photoluminescence (PL) and photoluminescence
excitation spectroscopy (PLE)  are performed at various
temperatures ranging from 10K to 300K. The origin of the
different PL lines is discussed and an extensive study of the
electron-phonon coupling is performed, for the first time to our
knowledge in this kind of perovskite molecular crystal. A
Longitudinal optical (LO) phonon energy of 14 meV and an
electron-phonon coupling more than one order of magnitude higher
than in GaAs quantum wells, have been estimated. PL and PLE
experiments suggest that the PL signal at low temperature has to
be interpreted in the context of a polaron model.

When a  solution of $C_{6}H_{5}C_{2}H_{4}-NH_{3}I$ and PbI$_{2}$
dissolved in stoechiometric amounts in DMF (N,N-DiMethylFormamide)
is deposited by spin-coating on a quartz substrate, a
self-organization occurs, leading to the formation of a molecular
crystal, consisting in an alternance of  organic layers and
inorganic layers (figure 1 a)). The self-organization occurs for
thicknesses from 3 nm to 100 nm. The thickness of the perovskite
layers can be controlled  by changing the concentration of the
solution, the speed and the acceleration of the spin-coater. AFM
(Atomic Force Microscopy) measurements show that a high quality
polycristalline film is obtained, with a surface rugosity as
small as 2 nm. X-ray diffraction has been used to characterize
the ordering within the self-organized perovskite layers under
study. Figure 1 shows an example of rocking curve measured on  a
50 nm PEPI thin layer (obtained by spin-coating a solution of 10
wt\% $C_{6}H_{5}C_{2}H_{4}-NH_{3}I$ and PbI$_{2}$ dissolved in
stoechiometric amounts in DMF). The observation of numerous
diffraction orders proves the high crystallinity of the thin
layer and the very good periodicity of the stacking. Each
satellite peak being regarded as a diffraction plane, it is seen
that we can observe diffraction planes from (002) to (0022)
beyond $2\theta = 60^{\mathrm{o}}$. As a consequence, a period of
16.4~\AA~can be accurately estimated. The inorganic layers, which
consist of PbI$_{6}^{2-}$ octahedrons, having a lattice parameter
of 6.4 \AA, we deduce that the organic part has an extension of
10~\AA~along the growth direction. Identical results have been
obtained from X-ray diffraction spectra on samples of different
thicknesses.

Because of the very large energy difference between the band gap
of the inorganic layer and the HOMO-LUMO of the organic layer, a
multi quantum well electronic structure is formed in these
self-organized perovskite structure with a strong 2D quantum
confinement: the quantum wells are formed by the PbI$_{6}^{2-}$
octahedrons monolayers and the barriers by the organic
alkylammonium layers. Moreover, the barrier and the well of these
natural QW have very different dielectric constants
($\epsilon_{barrier}\sim$ 2.1; $\epsilon_{well}\sim$
6.1)\cite{Ishihara90}. Thanks to this high contrast in dielectric
constants, the Coulomb interaction in the QW is hardly screened
by the presence of the barrier: it is the well known dielectric
confinement effect \cite{Keldysh79,Kumagai89}. Therefore, the
electron-hole interaction within the exciton is strengthened,
resulting in huge oscillator strengths and very large exciton
binding energies (of a few hundred of meV).

\begin{figure}
 \begin{center}
\includegraphics*[scale=.8]
{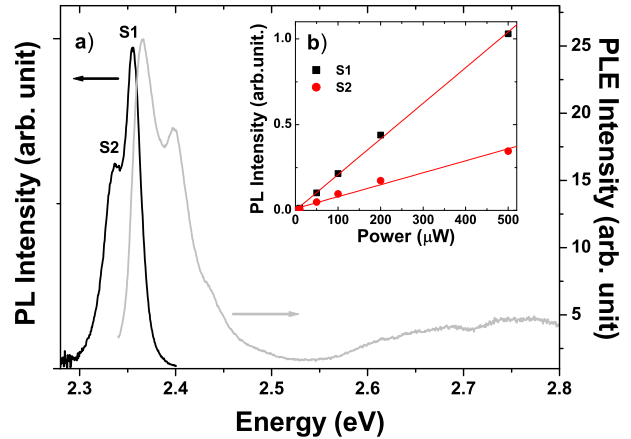} \caption[PL/PLE]{\textit{a) PL (black curve)
excited at 2.8 eV and PLE (grey curve) detected at the energy of S2
(2.337 eV) at 10K; b) Variation of the PL intensity of S1 and S2 as
a function of the pump power, at 2K.}} \label{fig2}
\end{center}
\end{figure}

PL and PLE experiments have been performed  on a PEPI layer of
thickness  $\sim 5$ nm, containing 3 QWs: this thin layer has been
obtained by spin-coating a solution of 1 wt\%
$C_{6}H_{5}C_{2}H_{4}-NH_{3}I$ and PbI$_{2}$ dissolved in
stoechiometric amounts in DMF. The excitation source is a
quartz-iode lamp spectrally filtered with a monochromator. The
image of the exit slit of the monochromator is formed on the
sample leading to a spot size of about 3 mm x 1 mm. PL experiments
as a function of the excitation power are performed using a laser
diode at 405 nm. The luminescence is then collected and dispersed
in a 1 m double-spectrometer and detected with a photomultiplier.
The sample can be placed either on the cold finger of a helium
closed cycle cryostat (sample temperature between 10K and 300K)
or in a bath of superfluid helium (2K).

Figure 2a exhibits the PL spectrum of the sample at 10K:  two
lines can be observed at 2.355 eV (S1) and 2.337 eV (S2). Similar
optical spectra were already reported in the literature but
interpreted in different ways
\cite{Fujisawa04a,Hong92a,Ishihara90}. Fujisawa et al attribute
the PL high energy line to the free excitons and the lower energy
lines to bound excitons \cite{Fujisawa04a}. Hong et al suggest
the presence of free excitons, bound excitons and of phonon
replica \cite{Hong92a} and Ishihara et al raise the possibility of
the existence of three exciton lines named A,B and C as observed
in GaN \cite{Ishihara90}. In order to get more information about
the two PL lines, PL spectra as a function of the excitation
power (at 2K) have been performed (see fig. 2b). Excitation
powers ranging between 3 $\mu W$ and 4 mW have been explored, but
results are presented only for excitation powers below 500 $\mu
W$ because the sample presents some photobleaching at higher
powers. As evidenced on figure 2b, the two PL lines have the same
behavior in this range of power. We can thus deduce that the
satellite line S2 has an intrinsic origin and should not be
attributed to excitons bound to defects (which would correspond
to a sublinear dependance on the excitation power) nor to
bi-excitons (surlinear behavior). Figure 2a shows also the PLE
spectrum recorded at 10K with a detection energy tuned in
resonance with the low energy PL emission line S2. The PLE
spectrum recorded with a detection energy tuned in resonance with
S1 is exactly similar: this again rules out an interpretation of
S1 in terms of bi-exciton. Moreover the independence of the PLE
spectra from the detection energy demonstrates that the two
emission lines are connected to the same excited states. The PLE
spectrum is composed of three lines at 2.365 eV, 2.398 eV and
2.436 eV, and of the 2D step around 2.565 eV which corresponds to
the absorption edge of the uncorrelated electron-hole pair
continuum. The energy difference between the first line and the
2D absorption step corresponds to an exciton binding energy of
200 meV, in good agreement with previous estimate deduced from
optical absorption spectroscopy \cite{Hong92a}. At 10 K, the
Stokes shift between the PLE lower energy line and the PL line S1
is 10 meV. The origin of the Stokes shift will be further
discussed in the following.

\begin{figure}
 \begin{center}
\includegraphics*[scale=.8]
{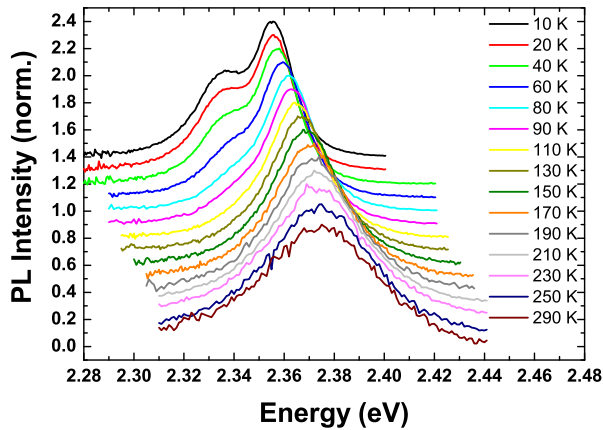} \caption[FWHM(T)]{\textit{Photoluminescence
spectra excited at 2.5 eV for temperatures ranging from 10 K to
295K.}} \label{fig3}
\end{center}
\end{figure}

We now investigate the optical properties of this perovskite
multi-layer sample for various temperatures. Figure 3 shows PL
spectra measured for temperatures ranging from 10K to 295K: the
energy of the high energy PL line increases with temperature. This
is in contrast with other perovskites, for instance
$(C_{n}H_{2n+1}-NH_{3})_{2}PbI_{4}$ for which the emission energy
has been reported to decrease with temperature \cite{Ishihara90}.
Our result shows the importance of the barrier nature for the
understanding of the perovskite electronic structure. Indeed,
depending on the exact composition of the barrier, the dielectric
constant difference $\epsilon_{well}-\epsilon_{barrier}$ between
the barrier and the QW may increase or decrease with temperature.
Moreover as detailed in ref. \cite{Hong92a}, the exciton energy
in perovskite QWs is mainly determined by
$\epsilon_{well}-\epsilon_{barrier}$ through the dielectric
confinement. As a result, changing the barrier composition may
lead to opposite temperature behavior of the exciton transition
energy. Additionally, the strain in the QW might be different
depending on the nature of the barrier~\cite{Zhang2009} and might
also influence the exciton energy dependance on temperature. A
detailed optical study of the temperature behaviour for several
perovksites is under progress and should further clarify the
exact role of the barrier composition.
\begin{figure}
 \begin{center}
\includegraphics*[scale=.5]
{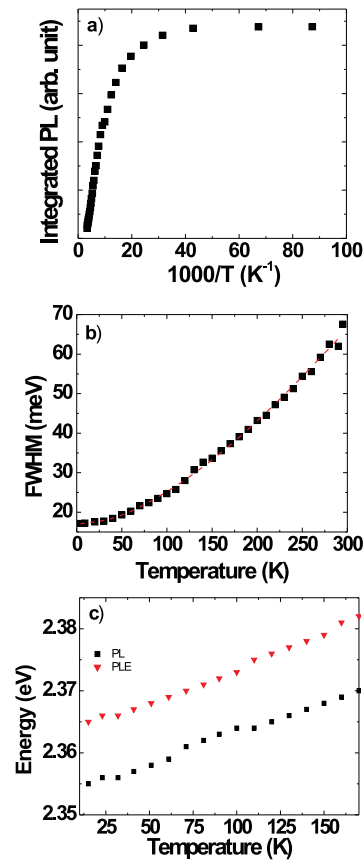} \caption[FWHM(T)]{\textit{a) Integrated PL
intensity of S1 as a function of 1000/T; b) Full Width at Half
maximum (FWHM) of the PL line S1 as a function of temperature; the
dashed line is the calculated values of the FWHM, following the
model explained in the text; c) Energy of the PL line S1 (squares),
and PLE lower energy line (triangles) as a function of
temperature.}} \label{fig4}
\end{center}
\end{figure}

Figure 4a summarizes the evolution with temperature of the
integrated  PL intensity measured on the S1 line. This emission
intensity remains stable between 10 K and 100 K and decreases by 2
order of magnitude between 100 K and 300 K, approximately as
exp(E$_{a}$/kT) with E$_{a}\sim$ 60 meV. This value is quite
different from the one measured by Hong et al \cite{Hong92a}: in
their paper, an activation energy of 230 meV is reported, quite
similar to the exciton binding energy. For that reason, the
decrease of the integrated PL intensity has been attributed  by
these authors to the partial ionization of the exciton.
Nevertheless, the activation energy we measure here is far from
the exciton binding energy. Performing similar measurements on
several samples of same thickness but prepared at different days,
E$_{a}$ is found to significantly vary from sample to sample
between 30 meV and 80 meV. The strong dispersion in the measured
value of E$_{a}$ demonstrates that the decrease of the integrated
PL intensity should be related to an extrinsic non radiative
mechanism.

As shown on fig. 4b, the measured full width at half maximum of
the PL line S1 continuously increases when increasing the sample
temperature between 10 K and 300 K. We attribute this broadening
to exciton interaction with phonons and describe it within a
phenomenological model, often used for inorganic quantum wells
\cite{Lee86}. The dashed line represents the fit of the
experimental data with the function:
\begin{equation}
\Gamma=\Gamma_{0}+a.T+\frac{\Gamma_{LO}}{exp(\frac{\hbar\omega_{LO}}{kT})-1}+\Gamma_{imp}.exp(\frac{-E_{a}}{kT})
\end{equation}
In this model, $\Gamma_{0}$ is the linewidth at 0 K, $a.T$
represents the scattering of excitons with acoustic phonons. The
third term corresponds to the interaction of excitons with optical
phonons, where $\Gamma_{LO}$ is related to the strength of
exciton-phonon coupling and $\hbar\omega_{LO}$ is the optical
phonon energy. The last term corresponds to the exciton
scattering on impurities with an activation energy $E_{a}= 60$ meV
deduced from temperature dependant measurements. Some assumptions
used in this model can be discussed in the case of
organic-inorganic perovskite quantum wells. First, the model is
valid within the framework of the envelope function \cite{Lee86},
which requires that the variation of the excitonic wave function
is smooth on several unit cells. In the perovskite molecular
crystal, the exciton Bohr radius has been estimated to be 1.7 nm
in the layers plane \cite{Hong92a}, that is to say 3 or 4 unit
cells. Therefore, the limit of the envelope function
approximation is reached in perovskite layers. Nevertheless,
notice that perovskite layers present several features of 2D QW
which are well described using the envelope function. Secondly,
 the thermal variation of optical phonons with temperature is neglected in our simple approach.
However in soft material, this assumption could be too strong. In
fact, the Van der Walls bounds which create the self-assembled QWs
are weak as compared to covalent bounds in inorganic QW systems.
As a consequence, a weak perturbation could strongly deform the
structure and affect the phonon energies. Anyway, using this
simple approach, we are able to give the first estimate of the
phonon energy and of exciton phonon-interaction in this new
system.

In order to extract physical values from the model, we have chosen
to fix the optical phonon energy $\hbar\omega_{LO}$ to 13.7 meV (the
value of optical phonons in PbI$_{2}$\cite{Skolnick78}). A very good
agreement between the experimental data and the theoretical model is
then obtained providing some numerical values for the parameters
$\Gamma_{0}$, $\Gamma_{LO}$ and $a$. A value of $\Gamma_{0}= 17\pm
1$ meV is deduced: it varies from a sample to another between 12 meV
and 17 meV, as expected if $\Gamma_{0}$ is mainly dominated by
inhomogeneous broadening. The parameter $a$ is equal to $0.03 \pm
0.01 meV.K^{-1}$. The uncertainty on this value is large because of
the large inhomogeneous linewidth at 0 K. Nevertheless, the exciton
interaction with acoustic phonons is found more than one order of
magnitude higher than in GaAs QWs \cite{Lee86}. The coupling
$\Gamma_{LO}$ with the optical phonons is estimated to $\sim$ 70 meV
which is also more than ten times higher than in inorganic quantum
wells \cite{Lee86}. These values don't depend on the sample, showing
the intrinsic nature of these measurements.

To finish the discussion of the perovskite optical spectroscopy,
we want to go further in the interpretation of the PL lines S1
and S2 and of the lines observed in the PLE spectrum. Figure 4c
shows that the Stokes shift between S1 and the lower energy PLE
line does not depend on temperature. In inorganic quantum wells,
the stokes shift is usually due to exciton localization in
potential minima induced by disorder. The exciton trapping
becomes less efficient as temperature is increased. As a result,
the stokes shift decreases with temperature and eventually
vanishes when the thermal energy kT overcomes the trapping
potential. The fact that in our Perovskite sample the 10 meV
Stokes Shift is still observed when kT is larger than 10 meV
(above 115 K) strongly suggests that the energy difference
between emission and absorption is not due to exciton
localization in disorder potential. It has been reported that lead
halides such as PbI$_{2}$, PbCl$_{2}$ or PbBr$_{2}$ exhibit a very
strong excitons-phonon coupling leading to the existence of
self-trapped excitons \cite{Plekhanov04}, and the Stokes Shift in
PbI$_{2}$ amounts to 7.4 $meV$ \cite{Goto96}. Moreover, note that
the independence of the Stokes shift with temperature can also be
observed in InGaN layers \cite{Kudo01} and has been interpreted as
an evidence of a polaron origin of the emission line. Since we
have shown previously that the exciton-phonon coupling is very
high in perovskite layers, the independence of the Stokes shift
with the temperature strongly suggests that the mean emission line
S1 has to be described in a polaron model. Since the energy
separation between S1 and S2 is 14 $meV$, which corresponds to
the LO phonon energy we have determined in PEPI, S2 may be a
phonon replica of the main line S1. Considering this context of
high exciton-phonon coupling, the three lines observed in the PLE
spectrum could also be related to phonon replica. Notice however
that the energy separation between the PLE lines is superior to
30 $meV$ (33 $meV$ between the lower energy line and the middle
one, 38 $meV$ between the middle line and the higher energy one).
This would mean that the vibration mode in the excited states is
different from that in the fundamental state, as it is often
observed in molecules or in molecular crystals where polaronic
effect occurs \cite{Jursenas1998}. Another interpretation of the
low energy PL line could be related to the electronic structure of
the conduction band: in PbI$_{2}$, the conduction band is three
times degenerate, the crystal field and the spin-orbit coupling
allowing the lift of the degeneracy \cite{Schluter74}. Therefore,
the different lines observed in PL and PLE spectra could come
from the different states of the conduction bands. Experiments as
a function of the polarization of the incident light are under
progress and should allow discriminating between the two
interpretations.

To conclude, we have performed optical spectroscopy
(photoluminescence and excitation of the photoluminescence) of
$(C_{6}H_{5}C_{2}H_{4}-NH_{3})_{2}PbI_{4}$ perovskite
self-organized layers.  For the first time an estimate of the
optical phonon energy is obtained as well as of the
exciton-phonon interaction. These results could be further
confirmed using Raman spectroscopy and theoretical calculations
of the Perovskite phonon structure. The exciton-phonon
interaction is found to be more than one order of magnitude
higher than in GaAs QWs, both for acoustic and optical phonons.
This leads to an interpretation of the PL emission lines within
the framework of a polaron model.  The very strong exciton-phonon
interaction in Perovskite QWs is very promising for the use of
this material as an active medium in polaritonic systems such as
microcavities operating in the strong coupling regime. It should
allow efficient polariton relaxation, a key requirement to obtain
polariton lasing.

Acknowledgments: Authors are grateful to M. Schott for helpful
discussions. This work has been supported by the ANR grant
"MICHRY", the C'Nano "R\'egion Ile de France" grant "MICROG" and
the RTRA grant "MOSKITO". The Laboratoire de Photonique Quantique
et Mol\'{e}culaire de l'Ecole Normale Sup\'{e}rieure de Cachan is
a Unit\'e Mixte de Recherche associ\'ee au CNRS (UMR8537).

\bibliography{prbpepi}

\begin{thebibliography}{37}
\expandafter\ifx\csname natexlab\endcsname\relax\def\natexlab#1{#1}\fi
\expandafter\ifx\csname bibnamefont\endcsname\relax
  \def\bibnamefont#1{#1}\fi
\expandafter\ifx\csname bibfnamefont\endcsname\relax
  \def\bibfnamefont#1{#1}\fi
\expandafter\ifx\csname citenamefont\endcsname\relax
  \def\citenamefont#1{#1}\fi
\expandafter\ifx\csname url\endcsname\relax
  \def\url#1{\texttt{#1}}\fi
\expandafter\ifx\csname urlprefix\endcsname\relax\def\urlprefix{URL }\fi
\providecommand{\bibinfo}[2]{#2}
\providecommand{\eprint}[2][]{\url{#2}}

\bibitem[{\citenamefont{{D.B. Mitzi, K. Chondroulis and C.R.
  Kagan}}(2001)}]{Mitzi01}
\bibinfo{author}{\bibnamefont{{D.B. Mitzi, K. Chondroulis and C.R. Kagan}}},
  \bibinfo{journal}{IBM J. RES. \& DEV.} \textbf{\bibinfo{volume}{45}},
  \bibinfo{pages}{29} (\bibinfo{year}{2001}).

\bibitem[{\citenamefont{{G.C. Papavassiliou and I.B.
  Koutselas}}(1995)}]{Papavassiliou95}
\bibinfo{author}{\bibnamefont{{G.C. Papavassiliou and I.B. Koutselas}}},
  \bibinfo{journal}{Synth. Met.} \textbf{\bibinfo{volume}{71}},
  \bibinfo{pages}{1713} (\bibinfo{year}{1995}).

\bibitem[{\citenamefont{{R. Parashkov, A. Br\'ehier, A. Georgiev, S. Bouchoule,
  X. Lafosse, J. S. Lauret, C. T. Nguyen, M. Leroux, and E.
  Deleporte}}(2007)}]{Parashkov2007}
\bibinfo{author}{\bibnamefont{{R. Parashkov, A. Br\'ehier, A. Georgiev, S.
  Bouchoule, X. Lafosse, J. S. Lauret, C. T. Nguyen, M. Leroux, and E.
  Deleporte}}}, \bibinfo{journal}{in Progress in Advanced Materials Research,
  edited by N. H. Voler Nova Science}  (\bibinfo{year}{2007}).

\bibitem[{\citenamefont{{S.Zhang, G.Lanty, J.S. Lauret, E.Deleporte, P.
  Audebert, and Laurent Galmiche.}}(2009)}]{Zhang2009}
\bibinfo{author}{\bibnamefont{{S.Zhang, G.Lanty, J.S. Lauret, E.Deleporte, P.
  Audebert, and Laurent Galmiche.}}}, \bibinfo{journal}{Acta Materiala}
  \textbf{\bibinfo{volume}{57}}, \bibinfo{pages}{3301} (\bibinfo{year}{2009}).

\bibitem[{\citenamefont{{G. Lanty, A. Br\'ehier, R. Parashkov, J.S. Lauret and
  E.Deleporte}}(2008)}]{Lanty2008a}
\bibinfo{author}{\bibnamefont{{G. Lanty, A. Br\'ehier, R. Parashkov, J.S.
  Lauret and E.Deleporte}}}, \bibinfo{journal}{New Journal of Physics}
  \textbf{\bibinfo{volume}{10}}, \bibinfo{pages}{065007}
  (\bibinfo{year}{2008}).

\bibitem[{\citenamefont{{G. Lanty, J. S. Lauret, E. Deleporte, S. Bouchoule,
  and X. Lafosse}}(2008)}]{Lanty2008b}
\bibinfo{author}{\bibnamefont{{G. Lanty, J. S. Lauret, E. Deleporte, S.
  Bouchoule, and X. Lafosse}}}, \bibinfo{journal}{App. Phys. Lett.}
  \textbf{\bibinfo{volume}{93}}, \bibinfo{pages}{081101}
  (\bibinfo{year}{2008}).

\bibitem[{\citenamefont{{A. Br\'ehier, R. Parashkov, J. S. Lauret, and E.
  Deleporte}}(2006)}]{Brehier2006}
\bibinfo{author}{\bibnamefont{{A. Br\'ehier, R. Parashkov, J. S. Lauret, and E.
  Deleporte}}}, \bibinfo{journal}{App. Phys. Lett.}
  \textbf{\bibinfo{volume}{89}}, \bibinfo{pages}{171110}
  (\bibinfo{year}{2006}).

\bibitem[{\citenamefont{{J. Wenus, R. Parashkov, S. Ceccarelli, A. Br\'ehier,
  J. S. Lauret, M. S. Skolnick, E. Deleporte, and D. G.
  Lidzey}}(2006)}]{Wenus2006}
\bibinfo{author}{\bibnamefont{{J. Wenus, R. Parashkov, S. Ceccarelli, A.
  Br\'ehier, J. S. Lauret, M. S. Skolnick, E. Deleporte, and D. G. Lidzey}}},
  \bibinfo{journal}{Phys. Rev. B} \textbf{\bibinfo{volume}{74}},
  \bibinfo{pages}{235212} (\bibinfo{year}{2006}).

\bibitem[{\citenamefont{{J.I. Fujisawa, and T.
  Ishihara}}(2004{\natexlab{a}})}]{Fujisawa04b}
\bibinfo{author}{\bibnamefont{{J.I. Fujisawa, and T. Ishihara}}},
  \bibinfo{journal}{Phys. Rev. B} \textbf{\bibinfo{volume}{70}},
  \bibinfo{pages}{113203} (\bibinfo{year}{2004}{\natexlab{a}}).

\bibitem[{\citenamefont{{J.I. Fujisawa, and T.
  Ishihara}}(2004{\natexlab{b}})}]{Fujisawa04a}
\bibinfo{author}{\bibnamefont{{J.I. Fujisawa, and T. Ishihara}}},
  \bibinfo{journal}{Phys. Rev. B} \textbf{\bibinfo{volume}{70}},
  \bibinfo{pages}{205330} (\bibinfo{year}{2004}{\natexlab{b}}).

\bibitem[{\citenamefont{{M. Shimizu, and J.I. Fujisawa}}(2004)}]{Shimizu04}
\bibinfo{author}{\bibnamefont{{M. Shimizu, and J.I. Fujisawa}}},
  \bibinfo{journal}{J. of Lumin.} \textbf{\bibinfo{volume}{108}},
  \bibinfo{pages}{189} (\bibinfo{year}{2004}).

\bibitem[{\citenamefont{{T. Kondo, T. Azuma, T. Yuasa, and R.
  Ito}}(1998)}]{Kondo98}
\bibinfo{author}{\bibnamefont{{T. Kondo, T. Azuma, T. Yuasa, and R. Ito}}},
  \bibinfo{journal}{Solid sate comm.} \textbf{\bibinfo{volume}{105}},
  \bibinfo{pages}{253} (\bibinfo{year}{1998}).

\bibitem[{\citenamefont{{K. Tanaka, F. Sano, T. Takahashi, T. Kondo, R. Ito,
  and K. Ema}}(2002)}]{Tanaka02}
\bibinfo{author}{\bibnamefont{{K. Tanaka, F. Sano, T. Takahashi, T. Kondo, R.
  Ito, and K. Ema}}}, \bibinfo{journal}{Solid sate comm.}
  \textbf{\bibinfo{volume}{122}}, \bibinfo{pages}{249} (\bibinfo{year}{2002}).

\bibitem[{\citenamefont{{T. Hattori, T. Taira, M. Era, T. Tsutsui and S.
  Saito}}(1996)}]{Hattori96}
\bibinfo{author}{\bibnamefont{{T. Hattori, T. Taira, M. Era, T. Tsutsui and S.
  Saito}}}, \bibinfo{journal}{Chem. Phys. Lett.}
  \textbf{\bibinfo{volume}{254}}, \bibinfo{pages}{103} (\bibinfo{year}{1996}).

\bibitem[{\citenamefont{{X. Hong, T. Ishihara and A.V.
  Nurmiko}}(1992{\natexlab{a}})}]{Hong92a}
\bibinfo{author}{\bibnamefont{{X. Hong, T. Ishihara and A.V. Nurmiko}}},
  \bibinfo{journal}{Phys. Rev. B} \textbf{\bibinfo{volume}{45}},
  \bibinfo{pages}{6961} (\bibinfo{year}{1992}{\natexlab{a}}).

\bibitem[{\citenamefont{{T. Ishihara, J. Takahashi and Takenari
  Goto}}(1989)}]{Ishihara89}
\bibinfo{author}{\bibnamefont{{T. Ishihara, J. Takahashi and Takenari Goto}}},
  \bibinfo{journal}{Solid State Comm.} \textbf{\bibinfo{volume}{69}},
  \bibinfo{pages}{933} (\bibinfo{year}{1989}).

\bibitem[{\citenamefont{{T. Ishihara, J. Takahashi and Takenari
  Goto}}(1990)}]{Ishihara90}
\bibinfo{author}{\bibnamefont{{T. Ishihara, J. Takahashi and Takenari Goto}}},
  \bibinfo{journal}{Phys. Rev. B} \textbf{\bibinfo{volume}{42}},
  \bibinfo{pages}{11099} (\bibinfo{year}{1990}).

\bibitem[{\citenamefont{{T. Ishihara, X. Hong, J. Ding and N.V.
  Nurmikko}}(1992)}]{Ishihara92}
\bibinfo{author}{\bibnamefont{{T. Ishihara, X. Hong, J. Ding and N.V.
  Nurmikko}}}, \bibinfo{journal}{Surface Science}
  \textbf{\bibinfo{volume}{267}}, \bibinfo{pages}{323} (\bibinfo{year}{1992}).

\bibitem[{\citenamefont{{C.Q. Xu, H. Sakakura, T. Kondo, S. Takeyama, N. Miura,
  Y. Takahashi, K. Kumata and R. Ito}}(1991)}]{Xu91a}
\bibinfo{author}{\bibnamefont{{C.Q. Xu, H. Sakakura, T. Kondo, S. Takeyama, N.
  Miura, Y. Takahashi, K. Kumata and R. Ito}}}, \bibinfo{journal}{Solid state
  comm} \textbf{\bibinfo{volume}{79}}, \bibinfo{pages}{249}
  (\bibinfo{year}{1991}).

\bibitem[{\citenamefont{{X. Hong, T. Ishihara and A.V.
  Nurmiko}}(1992{\natexlab{b}})}]{Hong92b}
\bibinfo{author}{\bibnamefont{{X. Hong, T. Ishihara and A.V. Nurmiko}}},
  \bibinfo{journal}{Solid state comm} \textbf{\bibinfo{volume}{84}},
  \bibinfo{pages}{657} (\bibinfo{year}{1992}{\natexlab{b}}).

\bibitem[{\citenamefont{{C.Q. Xu, S. Fukuta, H. Sakakura, T. Kondo,R. Ito, Y.
  Takahashi, and K. Kumata}}(1991)}]{Xu91b}
\bibinfo{author}{\bibnamefont{{C.Q. Xu, S. Fukuta, H. Sakakura, T. Kondo,R.
  Ito, Y. Takahashi, and K. Kumata}}}, \bibinfo{journal}{Solid state comm}
  \textbf{\bibinfo{volume}{77}}, \bibinfo{pages}{923} (\bibinfo{year}{1991}).

\bibitem[{\citenamefont{{T. Kataoaka, T. Kondo, R. Ito, S. Sasaki, K. Uchida,
  and N. Miura}}(1993)}]{Kataoka93}
\bibinfo{author}{\bibnamefont{{T. Kataoaka, T. Kondo, R. Ito, S. Sasaki, K.
  Uchida, and N. Miura}}}, \bibinfo{journal}{Phys. Rev. B}
  \textbf{\bibinfo{volume}{47}}, \bibinfo{pages}{2010} (\bibinfo{year}{1993}).

\bibitem[{\citenamefont{{E.A. Muljarov, S.G. Tikhodeev, N.A. Gippius and Teruya
  Ishihara}}(1995)}]{Muljarov95}
\bibinfo{author}{\bibnamefont{{E.A. Muljarov, S.G. Tikhodeev, N.A. Gippius and
  Teruya Ishihara}}}, \bibinfo{journal}{Phys. Rev. B}
  \textbf{\bibinfo{volume}{51}}, \bibinfo{pages}{14370} (\bibinfo{year}{1995}).

\bibitem[{\citenamefont{{M. Era, S. Morimoto, T. Tsutsui, and S.
  Saito}}(1994)}]{Era94}
\bibinfo{author}{\bibnamefont{{M. Era, S. Morimoto, T. Tsutsui, and S.
  Saito}}}, \bibinfo{journal}{App. Phys. Lett.} \textbf{\bibinfo{volume}{65}},
  \bibinfo{pages}{676} (\bibinfo{year}{1994}).

\bibitem[{\citenamefont{{T. Fujita, Y. Sato, T. Kuitani, and T.
  Ishihara}}(1998)}]{Fujita98}
\bibinfo{author}{\bibnamefont{{T. Fujita, Y. Sato, T. Kuitani, and T.
  Ishihara}}}, \bibinfo{journal}{Phys. Rev. B} \textbf{\bibinfo{volume}{57}},
  \bibinfo{pages}{12428} (\bibinfo{year}{1998}).

\bibitem[{\citenamefont{{A. Kavokin and G. Malpuech}}(2003)}]{Livrekavokin}
\bibinfo{author}{\bibnamefont{{A. Kavokin and G. Malpuech}}},
  \emph{\bibinfo{title}{Cavity Polaritons}} (\bibinfo{publisher}{Elsevier,
  Amsterdam}, \bibinfo{year}{2003}).

\bibitem[{\citenamefont{{D. Bajoni, P. Senellart,E. Wertz, I. Sagnes, A. Miard,
  A. Lemaitre, and J. Bloch}}(2008)}]{bajoniPRL}
\bibinfo{author}{\bibnamefont{{D. Bajoni, P. Senellart,E. Wertz, I. Sagnes, A.
  Miard, A. Lemaitre, and J. Bloch}}}, \bibinfo{journal}{Phys. Rev. Lett.}
  \textbf{\bibinfo{volume}{100}}, \bibinfo{eid}{047401} (\bibinfo{year}{2008}).

\bibitem[{\citenamefont{{S. Christopoulos and G. Baldassarri, H. von
  Hogersthal, A. J. D. Grundy, P. G. Lagoudakis, A. V. Kavokin, J. J. Baumberg,
  G. Christmann, R. Butte, E. Feltin, J.-F. Carlin, and N.
  Grandjean}}(2007)}]{christopoulos}
\bibinfo{author}{\bibnamefont{{S. Christopoulos and G. Baldassarri, H. von
  Hogersthal, A. J. D. Grundy, P. G. Lagoudakis, A. V. Kavokin, J. J. Baumberg,
  G. Christmann, R. Butte, E. Feltin, J.-F. Carlin, and N. Grandjean}}},
  \bibinfo{journal}{Phys. Rev. Lett.} \textbf{\bibinfo{volume}{98}},
  \bibinfo{eid}{126405} (\bibinfo{year}{2007}).

\bibitem[{\citenamefont{{L. V. Keldysh}}(1979)}]{Keldysh79}
\bibinfo{author}{\bibnamefont{{L. V. Keldysh}}}, \bibinfo{journal}{JETP Lett.}
  \textbf{\bibinfo{volume}{29}}, \bibinfo{pages}{658} (\bibinfo{year}{1979}).

\bibitem[{\citenamefont{{M. Kumagai, and T. Takagahara}}(1989)}]{Kumagai89}
\bibinfo{author}{\bibnamefont{{M. Kumagai, and T. Takagahara}}},
  \bibinfo{journal}{Phys. Rev. B} \textbf{\bibinfo{volume}{40}},
  \bibinfo{pages}{12359} (\bibinfo{year}{1989}).

\bibitem[{\citenamefont{{J. Lee, E. S. Koteles, and M.O.
  Vassel}}(1986)}]{Lee86}
\bibinfo{author}{\bibnamefont{{J. Lee, E. S. Koteles, and M.O. Vassel}}},
  \bibinfo{journal}{Phys. Rev. B} \textbf{\bibinfo{volume}{33}},
  \bibinfo{pages}{5512} (\bibinfo{year}{1986}).

\bibitem[{\citenamefont{{M.S. Skolnick, and D. Bimberg}}(1978)}]{Skolnick78}
\bibinfo{author}{\bibnamefont{{M.S. Skolnick, and D. Bimberg}}},
  \bibinfo{journal}{Phys. Rev. B} \textbf{\bibinfo{volume}{18}},
  \bibinfo{pages}{7080} (\bibinfo{year}{1978}).

\bibitem[{\citenamefont{{V.G. Plekhanov}}(2004)}]{Plekhanov04}
\bibinfo{author}{\bibnamefont{{V.G. Plekhanov}}}, \bibinfo{journal}{Progress in
  Materials Science} \textbf{\bibinfo{volume}{49}}, \bibinfo{pages}{787}
  (\bibinfo{year}{2004}).

\bibitem[{\citenamefont{{T. Goto, and S. Saito}}(1996)}]{Goto96}
\bibinfo{author}{\bibnamefont{{T. Goto, and S. Saito}}}, \bibinfo{journal}{J.
  Of Lumin.} \textbf{\bibinfo{volume}{70}}, \bibinfo{pages}{435}
  (\bibinfo{year}{1996}).

\bibitem[{\citenamefont{{H. Kudo, K. Murakami, H. Ishibashi, R. Zheng, Y.
  Yamada and T. Taguchi}}(2001)}]{Kudo01}
\bibinfo{author}{\bibnamefont{{H. Kudo, K. Murakami, H. Ishibashi, R. Zheng, Y.
  Yamada and T. Taguchi}}}, \bibinfo{journal}{Phys. Stat. Sol. (b)}
  \textbf{\bibinfo{volume}{228}}, \bibinfo{pages}{55} (\bibinfo{year}{2001}).

\bibitem[{\citenamefont{{S. Jursenas, A. Gruodis, G. Kodis, M. Chachisvilis, V.
  Gulbinas, E.A. Silinsh and L. Valkunas}}(1998)}]{Jursenas1998}
\bibinfo{author}{\bibnamefont{{S. Jursenas, A. Gruodis, G. Kodis, M.
  Chachisvilis, V. Gulbinas, E.A. Silinsh and L. Valkunas}}},
  \bibinfo{journal}{J. Phys. Chem. B} \textbf{\bibinfo{volume}{102}},
  \bibinfo{pages}{1086} (\bibinfo{year}{1998}).

\bibitem[{\citenamefont{{I. Ch. Schl\"{u}ter and M.
  Schl\"{u}ter}}(1974)}]{Schluter74}
\bibinfo{author}{\bibnamefont{{I. Ch. Schl\"{u}ter and M. Schl\"{u}ter}}},
  \bibinfo{journal}{Phys. Rev. B} \textbf{\bibinfo{volume}{9}},
  \bibinfo{pages}{1652} (\bibinfo{year}{1974}).

\end{thebibliography}

\end{document}